\documentclass[fleqn,usenatbib]{mnras}

\usepackage{newtxtext,newtxmath}

\usepackage[T1]{fontenc}

\DeclareRobustCommand{\VAN}[3]{#2}
\let\VANthebibliography\thebibliography
\def\thebibliography{\DeclareRobustCommand{\VAN}[3]{##3}\VANthebibliography}

\usepackage{graphicx}	
\usepackage{amsmath}	

\usepackage{relsize}
\usepackage{verbatim}

\usepackage[normalem]{ulem}
\usepackage{xcolor}
\usepackage{soul}
\setstcolor{red}

\title[Black Hole Images with Magnetic Tilts]{Observational Signatures of Black Hole Accretion: Rotating vs. Spherical Flows with Tilted Magnetic Fields}

\usepackage{CJK}
\author[H. Jia et al.]{
He Jia (\begin{CJK*}{UTF8}{gbsn}贾赫\end{CJK*})$^{1}$\thanks{E-mail: hejia@princeton.edu},
Christopher J. White$^{1}$,
Eliot Quataert$^{1}$
and Sean M. Ressler$^{2}$
\\
$^{1}$Department of Astrophysical Sciences, Princeton University, Princeton, NJ 08544, USA\\
$^{2}$Kavli Institute for Theoretical Physics, University of California Santa Barbara, Santa Barbara, CA 93107, USA
}

\date{Accepted XXX. Received YYY; in original form ZZZ}

\pubyear{2015}

\begin{document}
\label{firstpage}
\pagerange{\pageref{firstpage}--\pageref{lastpage}}
\maketitle

\begin{abstract}
We study the observational signatures of  magnetically arrested black hole accretion with non-rotating inflow onto a rotating black hole; we consider a range of angles between the black hole spin and the initial magnetic field orientation.  We compare the results of our General Relativistic Magneto-Hydrodynamic simulations to more commonly used rotating initial conditions  and to the Event Horizon Telescope (EHT)  observations of M87.   We find that the mm intensity images, polarization images, and synchrotron emission spectra are very similar among the different simulations when post-processed with the same electron temperature model; observational differences due to different electron temperature models are significantly larger than those due to the different realizations of magnetically arrested accretion.   
The orientation of the mm synchrotron polarization is particularly insensitive to the initial magnetic field orientation, the electron temperature model, and the rotation of the inflowing plasma.
The largest difference among the simulations with different initial rotation and magnetic tilt is in the strength and stability of the jet; spherical inflow leads to kink-unstable jets.  We discuss the implications of our results for current and future EHT observations and for theoretical models of event-horizon-scale black hole accretion.

\end{abstract}

\begin{keywords}
black hole physics -- accretion, accretion discs -- relativistic processes -- methods: numerical
\end{keywords}



\section{Introduction}

Using 230 GHz observations in April 2017, \citet{2019ApJ...875L...1E,2019ApJ...875L...4E}  presented the first images of the central supermassive black hole in M87. These were followed by the quasi-simultaneous broad-band M87 spectrum data \citep{2021ApJ...911L..11E}, as well as the recently released polarization measurements \citep{2021ApJ...910L..12E,2021ApJ...910L..13E}. These results not only improve our understanding of black hole astrophysics, but also open a new way to quantitatively study strong-field general relativity \citep[e.g.][]{Gralla2020}.

Traditionally, the process of black hole accretion was studied with semi-analytic models \citep[e.g.][]{1973A&A....24..337S,1977MNRAS.179..433B,1994ApJ...428L..13N,2000ApJ...539..798N}. To achieve higher complexity and capture more realistic physical processes, however, one needs to switch to numerical methods. At this time, most numerical work on black hole accretion has centered on General Relativistic Magneto-Hydrodynamic (GRMHD) simulations, which are commonly initialized with a torus of plasma orbiting the central black hole \citep{Fishbone1976,2003ApJ...589..444G,2013A&A...559A.116P,2017ComAC...4....1P}. A dynamically weak magnetic field, typically aligned with the black hole spin, seeds the magneto-rotational instability \citep[MRI,][]{1991ApJ...376..214B} and drives accretion. Such dynamical models, combined with post-processing ray tracing calculations of the synchrotron emission \citep{2009ApJ...696.1616D,2016MNRAS.462..115D,2018MNRAS.475...43M}, can produce asymmetric black hole images that qualitatively match the EHT observations \citep{2019ApJ...875L...5E}.

Despite the success to date, it should be emphasized that there are considerable uncertainties in GRMHD simulations of black hole accretion, with a large portion of the possible parameter space still unexplored.   For example, current theoretical models are broadly categorized as either Magnetically Arrested Disks \citep[MAD,][]{2003ApJ...592.1042I,2003PASJ...55L..69N} or Standard And Normal Evolution \citep[SANE,][]{2012MNRAS.426.3241N} depending on the magnitude of the magnetic flux threading the black hole. This quantity cannot, however, be directly measured with observations so far. Secondly, the exact value of the black hole spin $a$ in any particular system is uncertain. In addition, due to uncertainty in the  physics of particle heating in the near-horizon plasma, the relation between the proton and electron temperature is still unclear in systems like M87 \citep[e.g.][]{2019PhRvL.122e5101Z}, leading to a substantial uncertainty when computing synchrotron emission. Moreover, while it is typically assumed that the black hole spin, initial magnetic field and gas angular momentum are all aligned, this does not need to be the case; indeed, the tilted case(s) are an active research frontier \citep{2020ApJ...894...14W,2020MNRAS.499..362C,2021MNRAS.504.6076R}. These uncertainties need to be addressed before we can precisely constrain black hole accretion models and fundamental physics with astronomical observations.

In this paper, we study the observational signatures of tilted magnetic fields in black hole accretion: since there is no {\em a priori} reason to think that the  magnetic field in the inflowing plasma must necessarily be aligned with the black hole spin axis, it is more plausible to treat the tilt of the magnetic field  as a free parameter. This idea is implemented in the simulations presented in \citet{2021MNRAS.504.6076R}, where the initial angular momentum of the gas is set to zero so as to disentangle the effects of tilted magnetic field and tilted gas angular momentum \citep{2020ApJ...894...14W,2020MNRAS.499..362C}.

The remainder of this paper is organized as follows. Section \ref{sec:method} outlines the numerical methods, while section \ref{sec:flow} compares the fluid properties of accretion flows with different magnetic tilt and different amounts of initial gas rotation. We then present our results for intensity images in Section \ref{sec:intensity}, spectra in Section \ref{sec:spectra}, and polarimetric images in Section \ref{sec:polar}.  Throughout we compare to the EHT results on M87. We conclude in Section \ref{sec:discussion}.

\section{Methodology}

\label{sec:method}

We use the \textit{spherical} MAD simulations presented in \citet{2021MNRAS.504.6076R}, which employed \texttt{Athena++} \citep{2016ApJS..225...22W, 2020ApJS..249....4S} and were initialized with uniform, zero angular momentum gas and uniform magnetic fields with four different tilts ($0^{\circ}$, $30^{\circ}$, $60^{\circ}$ and $90^{\circ}$, which are labeled as Bz, B30, B60 and Bx respectively throughout this paper) with respect to the $a=0.9375$ black hole spin direction.  The GRMHD equations were evolved in Cartesian Kerr--Schild coordinates \citep{1963PhRvL..11..237K} extending at least $1{,}600$ gravitational radii $r_\mathrm{g} = G M / c^2$ in all directions, with nested static mesh refinement such that the resolution is comparable to or better than $192^3$ in logarithmic spherical Kerr--Schild coordinates. Note that by \textit{spherical} we mean the initial state of the ambient gas is spherically symmetric and has no angular momentum, but as the system evolves the gas will of course gain angular momentum through frame dragging of the central black hole and magnetic torques.

In all four cases, the gas is initially isothermal everywhere exterior to $6 \,r_g$ with a Bondi radius $r_\mathrm{Bondi} = 2 r_\mathrm{g} c^2 / c_\mathrm{s}^2 = 200\,r_\mathrm{g}$, where $c_\mathrm{s}$ is the sound speed of the gas, which has adiabatic index $\gamma = 5 / 3$. The simulations are run for $20{,}000$ gravitational times $r_\mathrm{g} / c$, corresponding to approximately $10$ free-fall times at the Bondi radius, with only data from times $t > 6{,}000\,r_\mathrm{g} / c$ used in this analysis. As shown in \citet{2021MNRAS.504.6076R}, these simulations attain a self-similar radial profile out to about $100\,r_\mathrm{g}$.

For comparison, we also use a standard $a=0.9375$ \textit{rotating} MAD torus simulation (labeled as MAD  hereafter).  The initial constant angular momentum \citet{Fishbone1976} torus has an inner radius of $r_{\rm in} = 20\,r_{\rm g}$ and a pressure maximum of $41\,r_{\rm g}$. The initial magnetic field is set via the vector potential $A_\varphi \propto \max(q,0),$ with
\begin{equation}
    q = \frac{\rho}{\rho_{\rm max}} \left(\frac{r}{r_{\rm in}}\right)^3 \sin^3(\theta)\exp\left(-\frac{r}{400r_{\rm g}}\right) - 0.2,
\end{equation}
where $\rho$ is the fluid frame mass density, $\rho_{\rm max} = 1$, and the proportionality constant is set so that the maximum pressure in the torus divided by the maximum magnetic pressure in the torus is 100. Small perturbations are added to the torus pressure at 2\% level. The adiabatic index for this simulation is $\gamma = 13/9$.  The MAD torus simulation is run for $10^4 \,r_g/c$ and reaches approximate inflow equilibrium (based on constancy of $\dot M$) out to $\simeq 30 \,r_g$. While the fluid state beyond this and especially beyond the initial pressure maximum where the simulation still retains memory of the initial condition is not expected to be representative of what is found in Nature, our results -- even when considering very faint emission as in Sec.~\ref{sec:polar} -- are only negligibly affected by material beyond ${\sim}16\,r_g$.

To calculate the radiative properties of these simulations, we adopt a series of electron temperature post-processing models, with different fixed values of electron-proton temperature ratio $T_e / T_p$. The electron temperature $T_e$ can be computed from GRMHD fluid pressure $p$ and density $\rho$ with
\begin{equation}
    T_e = \frac{2}{1+T_p/T_e} T_{\mathrm{fluid}} = \frac{m_p p}{(1 + T_p / T_e) k_B \rho},
\end{equation}
where $T_{\mathrm{fluid}}= (T_e + T_p) / 2 = m_p p / (2 k_B \rho)$ is defined as the (total) fluid temperature. In this paper, we consider $T_e / T_{\mathrm{fluid}}$ = $1/16$, $1/8$, $1/4$, $1/2$, $1$ and $2$. Motivated by the heating of electrons in a collisionless plasma, it has been argued that the ratio $T_e / T_p$ may depend on the plasma $\beta$ \citep{1999ApJ...520..248Q,2010MNRAS.409L.104H,2019PhRvL.122e5101Z}. We also carried out a number of calculations with the $R_{\mathrm{high}}-R_{\mathrm{low}}$ model of \citet{2016A&A...586A..38M}, which includes such $\beta$ dependence, but found that the main points of this work about the differences (or lack thereof) between spherical and rotating MADs are very similar with the two types of electron temperature models. This is consistent with Fig.~3 of \citet{2019ApJ...875L...5E} where they show that different $R_{\mathrm{high}}$ models generate almost identical images for MAD simulations (note that the $R_{\mathrm{high}}$ = $1$ model is the same as $T_e / T_{\mathrm{fluid}}$ = $1$). For brevity, we only show the results for the constant $T_e/T_{\mathrm{fluid}}$ models in this paper.

Once the electron and proton temperatures are determined, radiation from various channels can be evaluated. We focus on synchrotron radiation from hot electrons, and use the \texttt{grtrans} \citep{2009ApJ...696.1616D,2016MNRAS.462..115D} ray tracing code to generate intensity and polarimetric images from the GRMHD simulation data. Since non-radiative GRMHD simulations are scale-free, one needs to specify the black hole mass $M$ and fluid density scale $[\rho]$ in the ray tracing calculation. To match the M87 observations, throughout this paper, we set $M=6.5 \times 10^9\,M_{\odot}$ and the distance to M87 as $D=16.8\,\mathrm{Mpc}$ \citep{2009ApJ...694..556B, 2010A&A...524A..71B, 2018ApJ...856..126C}. The camera is located at $r_0=100\,r_g$, with viewing angle $\theta_0=163^{\circ}$ \citep{2016A&A...595A..54M} and azimuthal angle $\phi_0=0^{\circ}$, and the images are rotated such that the approaching jet has a position angle of $288^{\circ}$ \citep{2018ApJ...855..128W}, pointing towards the right and slightly up. We also tried another (arbitrarily chosen) viewing angle of $\theta_0=135^{\circ}$, but found that the main conclusions of this paper are insensitive to the exact value of $\theta_0$ (so they should apply to systems other than M87). For each electron temperature model, we calibrate the fluid density scale $[\rho]$ such that the total time-averaged intensity flux $F_{\nu}=1.2\,\mathrm{Jy}$ at $\nu=230\,\mathrm{GHz}$ \citep{2019ApJ...875L...4E}. We apply a cutoff for $\sigma=B^2 / (4 \pi \rho c^2)$ and ignore the emissions from regions where $\sigma > 1$, which are not trustworthy due to numerical issues in GRMHD simulations.

\begin{figure}
	\includegraphics[width=\columnwidth]{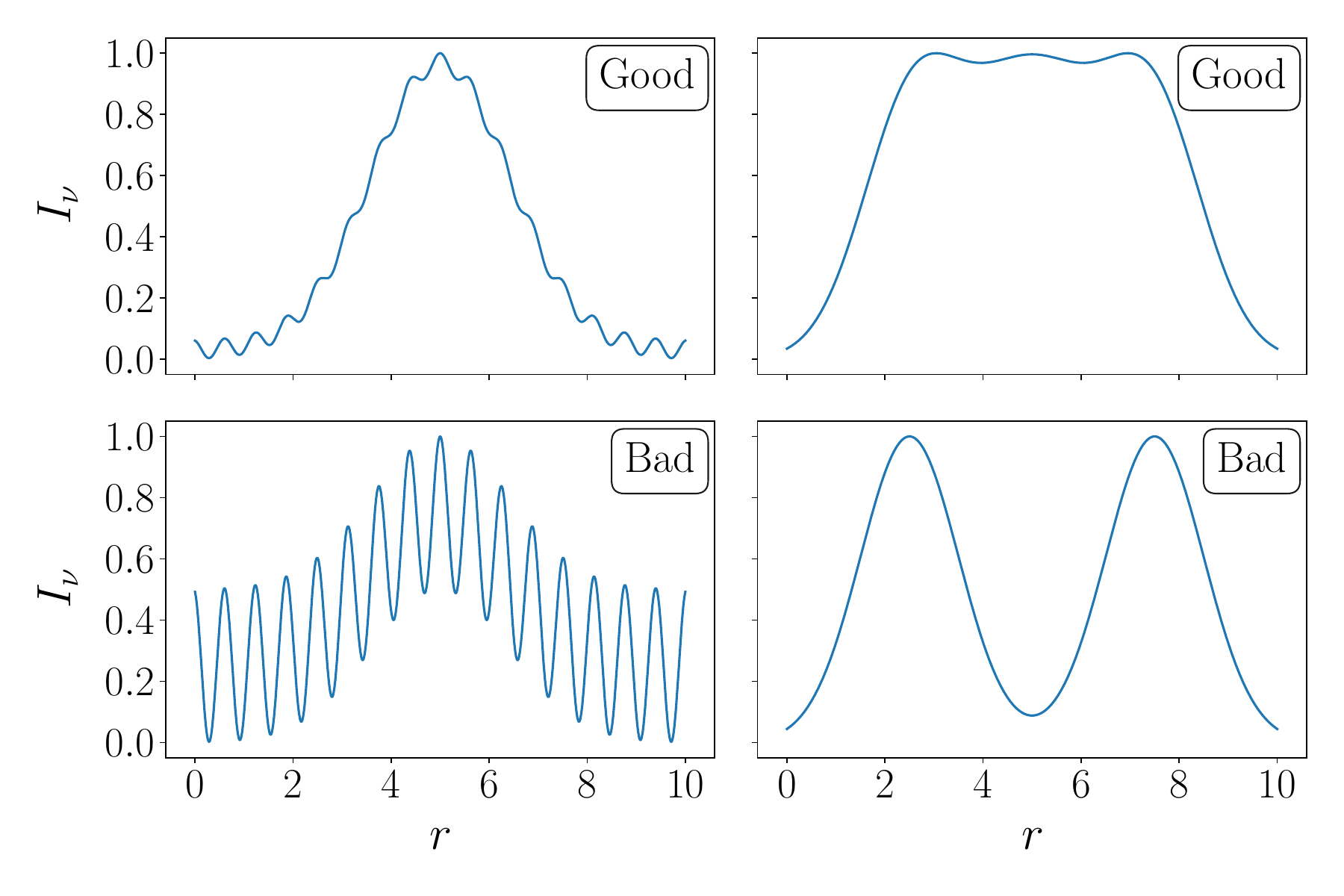}
    \caption{Toy examples of \textit{good} and \textit{bad} directions for intensity image analysis. In each panel, we plot the (blurred) 1-d slice of relative intensity as a function of the distance to the ring center in the image. We consider the small wiggles in the top panels as tolerable, but exclude the big fluctuations in the bottom panels from certain steps of our analysis to enhance robustness in defining the properties of the ring in our simulated images.}
    \label{fig:good-5-1}
\end{figure}

\begin{figure*}
	\includegraphics[width=\textwidth]{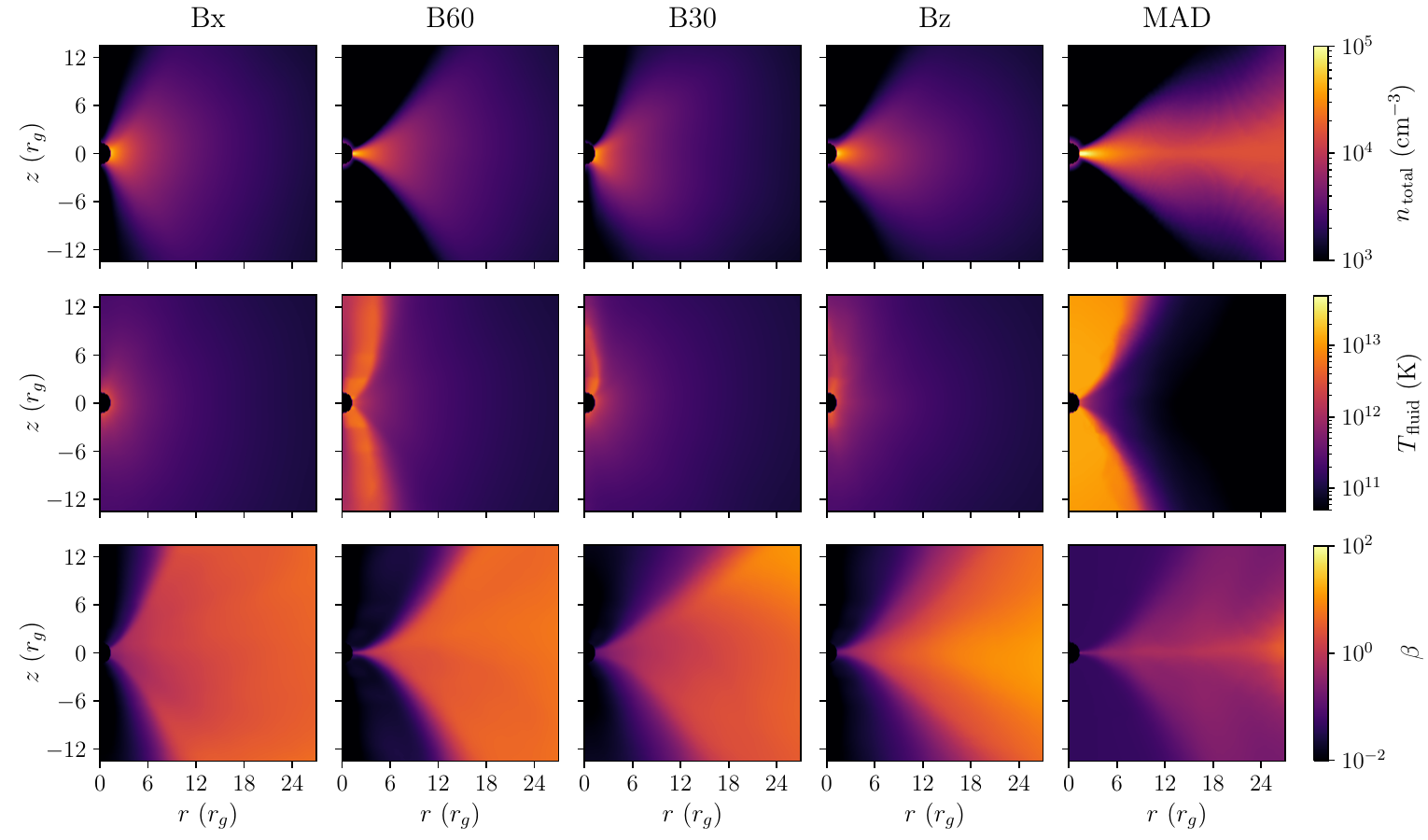}
	\caption{2-d plots of the density, temperature and plasma $\beta$, averaged over time and azimuthal angle, for the different simulations studied in this work. For $T_{\mathrm{fluid}}$ we take the density-weighted average $\left<n_{\rm total} T_{\mathrm{fluid}}\right> / \left<n_{\rm total}\right>$, while for $\beta$ we compute the harmonic mean $1/\left<1/\beta\right>$. The overall horizon scale fluid structure does not depend much on the tilt of initial magnetic field, and is similar for rotating and non-rotating initial conditions. The largest difference among the simulations is the stability of the jet propagation, shown in Fig.~\ref{fig:volume-rendering}.}
    \label{fig:nTbeta-0}
\end{figure*}

\begin{figure*}
	\centering
	\includegraphics[width=\textwidth]{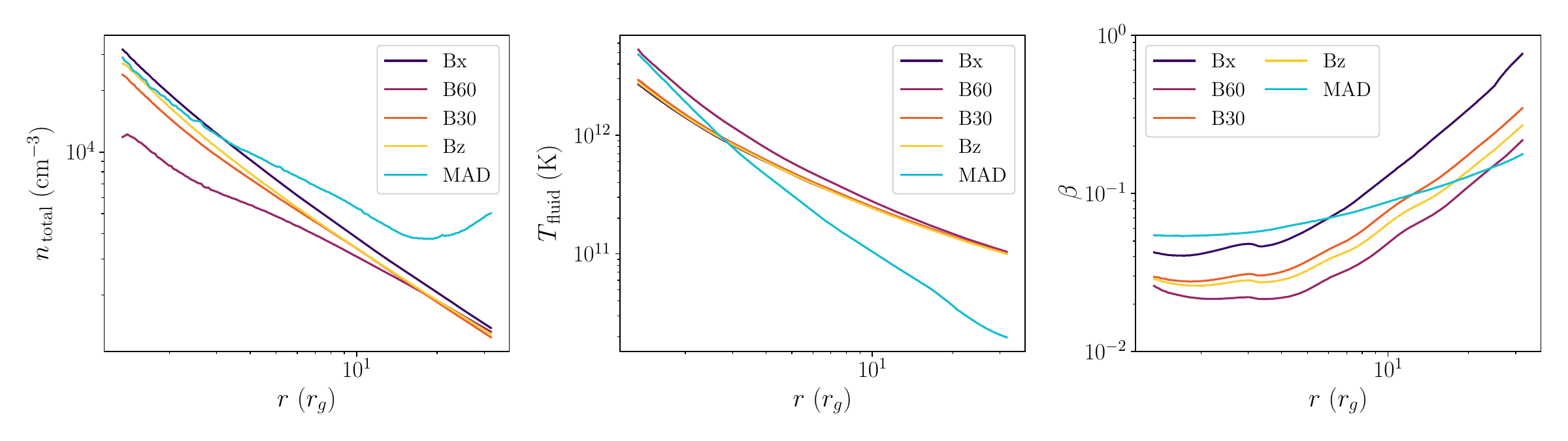}
	\caption{1-d plots of the fluid properties, same as Fig.~\ref{fig:nTbeta-0} but further averaged over the polar angle.}
    \label{fig:nTbeta-1}
\end{figure*}

\begin{figure*}
	\includegraphics[width=\textwidth]{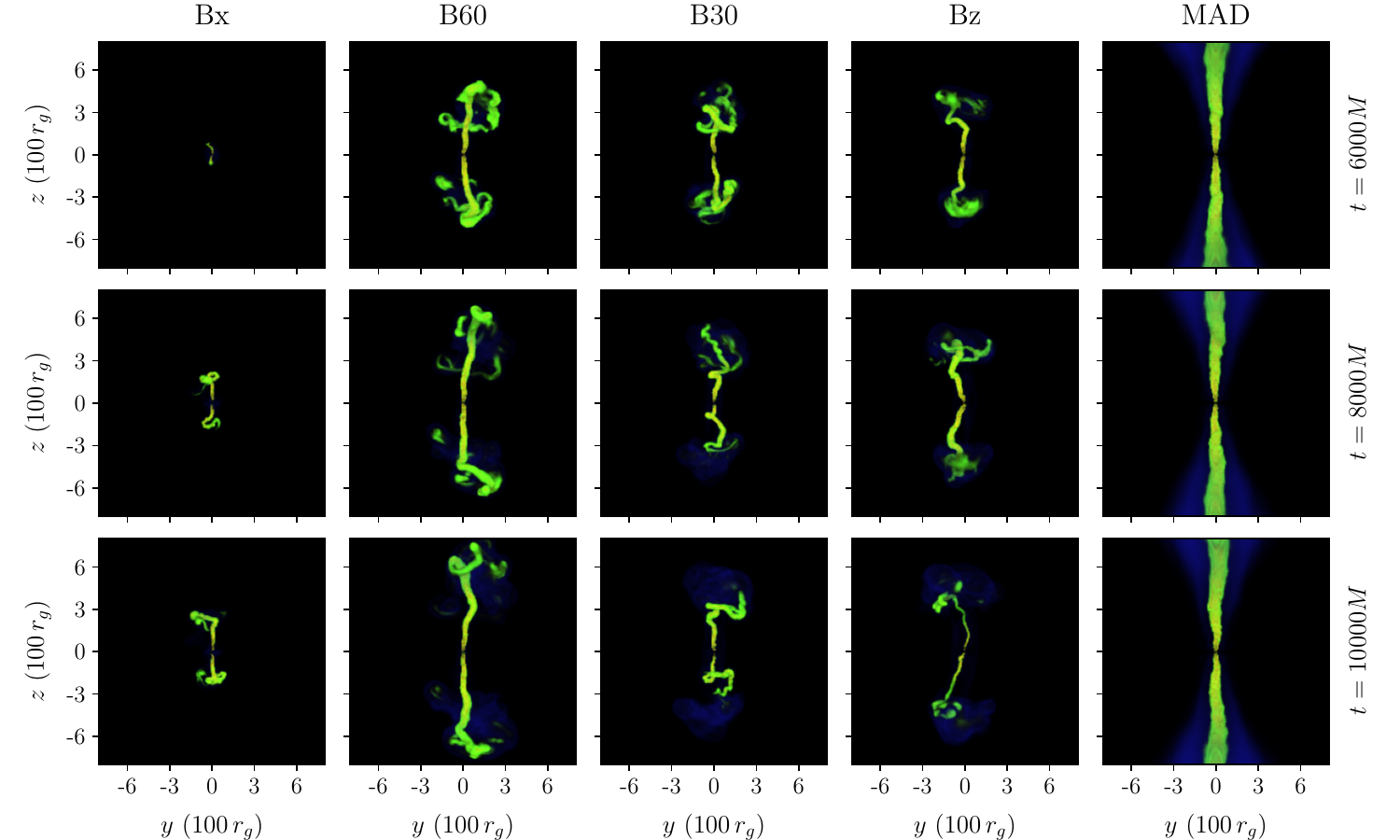}
	\caption{Volume rendering plots for the jets in the different simulations studied in this work, with warmer colors representing larger $\sigma = B^2 / (4 \pi \rho c^2)$. Roughly speaking, yellow structures have $\sigma \sim 20$, while for green regions $\sigma \sim 1$. For each simulation, we show the plots of three different snapshots. Generally, MAD simulations initialized with non-zero gas angular momentum have much more stable jets than the spherical simulations. Among the spherical cases, the jet is stronger in B60 and Bz, but all the spherical simulations show evidence of the kink instability that disrupts the jet propagation. Note that this figure shows the jet structure at the scale of hundreds of $r_g$, whereas only the central $\sim 10\, r_g$ region is directly relevant to EHT images. Movies containing all the snapshots for the five simulations are available at \url{https://youtube.com/playlist?list=PLYoKG8J1VGtFpyOJSDwQNqTe7rFJAOCdK} and also in the Supplementary Material.}
    \label{fig:volume-rendering}
\end{figure*}

To quantitatively compare the black hole images, we use a set of ring statistics as described below, similar to those used in EHT analysis \citep{2018ApJ...857...23C,2019ApJ...875L...4E,2021ApJ...910L..13E}. The main difference is that we introduce a distinction between \textit{good} and \textit{bad} directions to make the analysis more robust when the image does not have a well-defined ring structure.
\begin{enumerate}
    \item To locate the center of the ring, at each point $(x, y)$, we extract 1-d radial intensity profiles out to $60\,\mu$as (corresponding to $15.7\,r_g$ for M87), along $n_{\theta}=100$ different directions $\theta$ evenly spaced between $0^{\circ}$ and $360^{\circ}$. The choice of $60\,\mu$as should be enough as we find that the outer radius of the ring $r_{\mathrm{out}}$ (defined in step iv below) is at most about $40\,\mu$as for the cases considered in this paper. We linearly interpolate the intensity between the pixel grids.
    \item If, for example, the radial intensity profile is isotropic and multimodal like the bottom panels in Fig.~\ref{fig:good-5-1}, the image will be a series of concentric circles, inconsistent with the EHT observation data. Therefore, for a single well-defined ring, from the center the 1-d intensity profiles should be roughly unimodal along all the directions. To impose this constraint, we find the global peak of the radial profile for each direction, as well as all the \textit{high} local peaks (higher than 50\% of the global peak). If such high local peak exists, we require that between the innermost and outermost high peaks, the intensity should always be larger than 85\% times the lowest of the aforementioned high local peaks. Otherwise, the 1-d profile is considered to have non-trivial multi-modality, and we call the corresponding direction $\theta$ a \textit{bad} direction. Some toy examples of good and bad directions are given in Fig.~\ref{fig:good-5-1}. We do not simply require the 1-d profile of a good direction to be strictly unimodal, since for example the small ripples in the upper panels of Fig.~\ref{fig:good-5-1} should also be acceptable.
    \item For each \textit{good} direction, we define the ring radius as the location of the global intensity peak $r_{\mathrm{pk}}(\theta;x,y)$, and compute its mean $\Bar{r}_{\mathrm{pk}}(x,y)$ and standard deviation $\sigma_{\Bar{r}}(x,y)$ among all the good directions. Note that bad directions are excluded from $\Bar{r}_{\mathrm{pk}}(x,y)$ and $\sigma_{\Bar{r}}(x,y)$ computations. Suppose that $n_{\,\mathrm{bad}}(x,y)$ directions are bad, we optimize the following loss function, defined as the relative dispersion of the peak radius plus a penalty for bad directions,
    \begin{equation}
        f_{\,\mathrm{loss}}(x,y)=\sigma_{\Bar{r}}(x,y)/\Bar{r}_{\mathrm{pk}}(x,y)+5n_{\,\mathrm{bad}}(x,y)/n_{\theta},
    \end{equation}
    using the Nelder-Mead method \citep{10.1093/comjnl/7.4.308} to find the position of the ring center $(x_0,y_0)$. We will omit the argument of $(x_0,y_0)$ in the remainder of this section for simplicity.
    \item At the ring center $(x_0,y_0)$, we find $r_{\mathrm{pk}}(\theta)$, the global peak of the 1-d intensity profile; inner ring radius $r_{\mathrm{in}}(\theta)$, the smallest $r$ such that $I(r,\theta)=0.5I(r_{\mathrm{pk}}(\theta))$; and outer ring radius $r_{\mathrm{out}}(\theta)$, the largest $r$ such that $I(r,\theta)=0.5I(r_{\mathrm{pk}}(\theta))$. If the intensity at $r=0$ is already larger than $0.5I(r_{\mathrm{pk}}(\theta))$, we also consider this a bad direction. We average $r_{\mathrm{pk}}(\theta)$, $r_{\mathrm{in}}(\theta)$ and $r_{\mathrm{out}}(\theta)$ over all the good directions, and define the ring diameter $d=2r_{\mathrm{pk}}$ and ring width $w=r_{\mathrm{out}}-r_{\mathrm{in}}$.
    \item The ring orientation $\eta$ and degree of azimuthal asymmetry $A$ are defined via the first Fourier mode of the angular profile,
    \begin{align}
        \eta=\left< \mathrm{Arg} \left[ \int_0^{2\pi} I(\theta) e^{i\theta} d\theta \right] \right>_{r \in [r_{\mathrm{in}}, r_{\mathrm{out}}]},\
        \label{eq:eta}
    \end{align}
    \begin{equation}
        A=\left< \frac{ \left|\, \int_0^{2\pi} I(\theta) e^{i\theta}d\theta \,\right| }{\int_0^{2\pi} I(\theta) d\theta} \right>_{r \in [r_{\mathrm{in}}, r_{\mathrm{out}}]}.
        \label{eq:A}
    \end{equation}
    The fractional central brightness $f_C$ is defined as the ratio of the averaged intensity at the center to the averaged intensity around the ring,
    \begin{equation}
        f_C=\frac{\left< I(r,\theta) \right>_{\theta \in \left[ 0, 2\pi \right],\, r\in \left[ 0, 5\mu\mathrm{as} \right] }}{ \left< I(d/2, \theta) \right>_{\theta \in \left[ 0, 2\pi \right]} }.
        \label{eq:f_C}
    \end{equation}
    Note that the radial averaging in equations \ref{eq:eta}-\ref{eq:f_C} is weighted by $r$, so it represents the average over the 2-dim surface.
    \item To analyze the polarimetric properties, we follow \citet{palumbo2020discriminating} and \citet{2021ApJ...910L..13E} and compute the pixel-level image-averaged linear polarization fraction,
    \begin{equation}
        \left<|m|\right>=\frac{\mathlarger{\sum} _i \sqrt{\mathcal{Q}^2_i + \mathcal{U}^2_i}}{\mathlarger{\sum} _i \mathcal{I}_i},
        \label{eq:m}
    \end{equation}
    where we sum over all the pixels and time frames. We also define the $\beta_m$ coefficients in polar coordinates $(\rho,\phi)$ as
    \begin{equation}
        \beta_m=\frac{1}{I_{\mathrm{ann}}} \int_{\rho_{\min}}^{\rho_{\max}} \int_0^{2\pi} (\mathcal{Q} + i\mathcal{U}) e^{-im\phi} \rho d\phi d\rho,
        \label{eq:beta_m}
    \end{equation}
    where $I_{\mathrm{ann}}$ is the total intensity flux between $\rho_{\min}$ and $\rho_{\max}$. We take $m=2$, $\rho_{\min}=0$, and $\rho_{\max}$ to be large enough to contain the whole ring in the image.
\end{enumerate}

\section{Flow Properties at Horizon Scale}

\label{sec:flow}

\begin{figure*}
	\includegraphics[width=\textwidth]{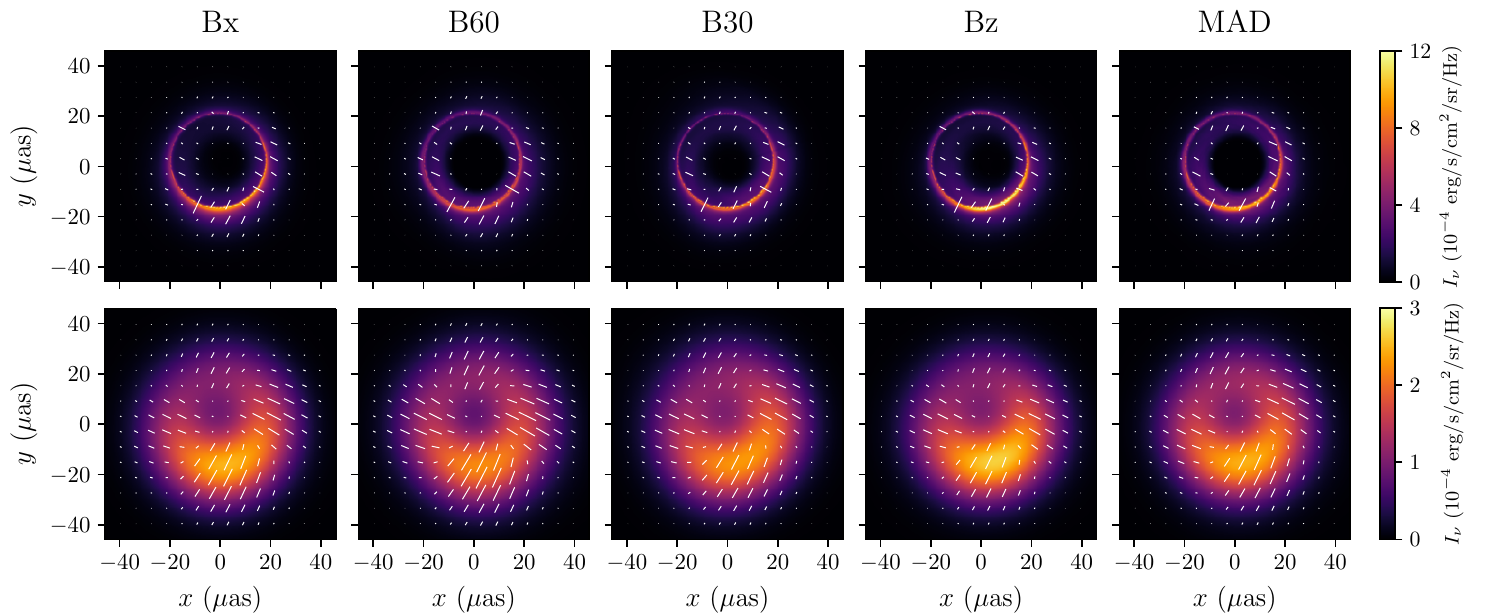}
    \caption{Black hole images from our various simulations with the $T_{\mathrm{e}}=T_{\mathrm{fluid}}/2$ electron temperature model, averaged over time. The tick length is proportional to the linearly polarized flux $\mathcal{P}=\sqrt{\mathcal{Q}^2+\mathcal{U}^2}$, while the tick direction stands for the direction of linear polarization. The bottom row shows images blurred with a 20$\,\mu$as Gaussian kernel to mimic finite EHT resolution. These images from different GRMHD simulations with the same electron temperature model are almost identical.}
    \label{fig:polar-img-0}
\end{figure*}

In Fig.~\ref{fig:nTbeta-0}, we compare the 2-d averaged number density $n_{\mathrm{total}}$, temperature $T_{\mathrm{fluid}}$ and plasma $\beta$ for the different simulations. We set the overall fluid density scale $[\rho]$ such that the time averaged flux $F_{\nu}=1.2\,$Jy for the baseline $T_e=0.5\,T_{\mathrm{fluid}}$ model. To put more emphasis on the region relevant for the emission, for $T_{\mathrm{fluid}}$ we weight the average by the number density $n_{\mathrm{total}}$, while for $\beta$ we show the harmonic mean, namely $1/\left<1/\beta\right>$. Here it is evident that B60 and MAD have stronger jets and more evacuated polar regions than other simulations. However, apart from this, we find that the fluid properties are overall similar among the different simulations at the horizon scale.

\begin{figure*}
	\includegraphics[width=15cm]{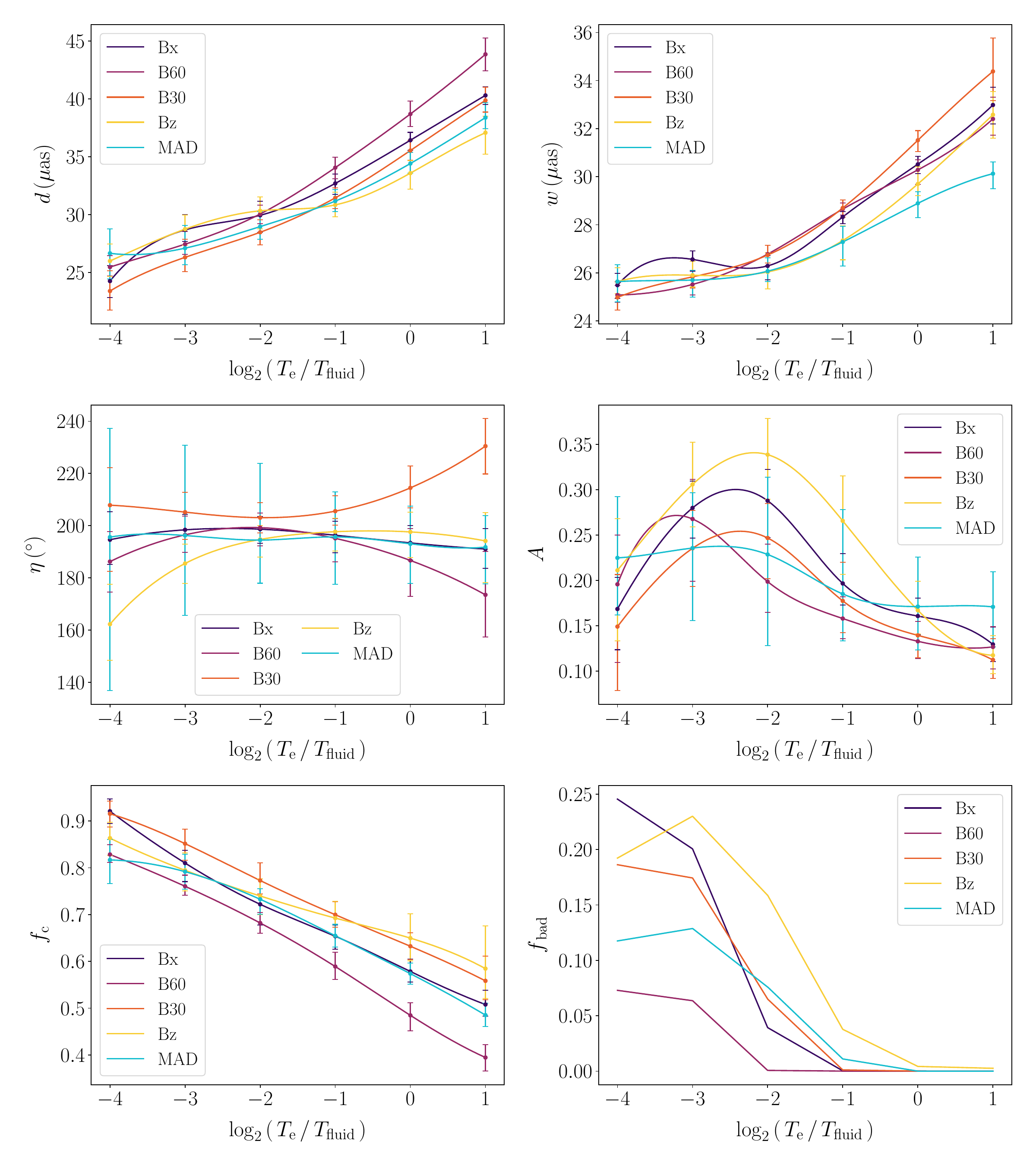}
    \caption{Black hole ring statistics for intensity images, including the ring diameter $d$, ring width $w$, orientation angle (east of north) $\eta$, asymmetry parameter $A$, fractional central brightness $f_C$ and the bad direction fraction $f_{\mathrm{bad}}$. The values are reported in terms of the 25\%, 50\% and 75\% percentiles over all the snapshots. Overall, the ring statistics are quite similar between the different GRMHD models.  In particular, differences due to the choice of the electron temperature model are much larger than those due to the choice of the GRMHD simulation.}
    \label{fig:ring-reports-0}
\end{figure*}

We further average over the polar angle to get the 1-d fluid property plots in Fig.~\ref{fig:nTbeta-1}. Since we weight the averaged $T_{\mathrm{fluid}}$ by density, it mainly reflects the temperature of the disk (instead of the jet) where most of the emission comes from in many models \citep{2019ApJ...875L...5E}. As we calibrate the time-averaged flux to the same value of $1.2\,$Jy for all the simulations, the synchrotron emissivity should be similar near the horizon. So for example, while B60 has a slightly smaller number density $n_{\mathrm{total}}$ just outside the horizon, this is compensated by a marginally larger temperature $T_{\mathrm{fluid}}$ and magnetic field strength. Again, the fluid properties are generally similar among these simulations. The modest differences in the B60 simulation relative to the other spherical simulations reflects the fact that the B60 simulation has a lower accretion rate (relative to Bondi), larger magnetic flux, and larger jet power than the other spherical simulations (see \citealt{2021MNRAS.504.6076R}).

In Fig.~\ref{fig:volume-rendering}, we use volume rendering \citep{lichtenbelt1998introduction} to visualize the jets in the different simulations. Widely used in e.g. Magnetic Resonance Imaging and Computed Tomography scans, volume rendering can display the 3-d structures better in 2-d images than simply doing a projection. We find that the jets in these simulations are quite different on larger scales. The MAD simulation initialized with non-zero angular momentum has the strongest jet, and is the only one in which the jet is still stable out to $10^3\,r_g$. Among the spherical simulations, the jet goes the farthest in the B60 simulation, out to around $800\,r_g$, while for Bx the jet is relatively feeble and even disappears in some snapshots. \citet{2021MNRAS.504.6076R} argued that this is due to the kink instability, likely triggered by the pressure confinement of the jet from the ambient polar gas. This low angular momentum polar material is not present in the rotating MAD initial conditions. 

\section{Intensity Images at 230 GHz}

\label{sec:intensity}

We generate ray tracing images with $256^2$ pixel resolution for all the snapshots in the spherical and standard MAD simulations, with constant electron-fluid temperature ratios $T_\mathrm{e}/T_{\mathrm{fluid}}=1/16,1/8,1/4,1/2,1$ and 2. We blur each image using a Gaussian kernel with its full-width half-maximum (FWHM) set to $20\,\mu$as to account for the EHT visibility measurement errors. The averaged images with $T_e/T_{\mathrm{fluid}}=1/2$ are shown in Fig.~\ref{fig:polar-img-0}, where the white ticks depicts the polarimetric patterns that will be discussed in details in Section \ref{sec:polar}. All of the images are remarkably similar to each other, with a prominent azimuthally asymmetric ring of emission.

Fig.~\ref{fig:ring-reports-0} shows the values of ring diameter $d$, ring width $w$, orientation angle (east of north) $\eta$, asymmetry parameter $A$, fractional central brightness $f_C$, as well as the bad direction fraction $f_{\mathrm{bad}}$ for our various dynamical and electron temperature models. These are measured using the blurred images (e.g. the bottom panels in Fig.~\ref{fig:polar-img-0}). Some of the ring statistics, like the ring width $w$, are highly sensitive to the size of blurring kernel, similar to the fact that \texttt{DIFMAP}, \texttt{eht-imaging} and \texttt{SMILI}, the three image reconstruction pipelines in EHT, give different ring width $w$ due to the difference in restoring beam size \citep{2019ApJ...875L...4E}. However, it is actually not clear how the visibility measurement errors would transform into the blurring of the images, especially when the visibility space is only sampled sparsely. Therefore, a more rigorous comparison to EHT data should be done in \textit{visibility} space. In this paper, since our main goal is to study the effects of different GRMHD models on black hole images rather than detailed comparison to EHT data, we stick to \textit{image} space for better intuition and interpretability.

For all the quantities except $f_{\mathrm{bad}}$, Fig.~\ref{fig:ring-reports-0} shows the median over all the snapshots, with error bars given by the 25\% and 75\% percentiles. We note that $f_{\mathrm{bad}}$ can be as large as 0.25 for lower electron temperature models, meaning that a non-trivial fraction of directions are \textit{bad}. If we instead blur the images with a 15$\,\mu$as FWHM Gaussian kernel, we find that for all the models $f_{\mathrm{bad}} \lesssim 0.02$, namely the ring structure is almost always well-defined. Nonetheless, the choice between 15$\,\mu$as and 20$\,\mu$as kernels does not significantly affect how the ring statistics depend on different GRMHD and electron temperature models, so we stick with 20$\,\mu$as for consistency with EHT analysis \citep{2019ApJ...875L...5E}.

\begin{figure}
	\includegraphics[width=\columnwidth]{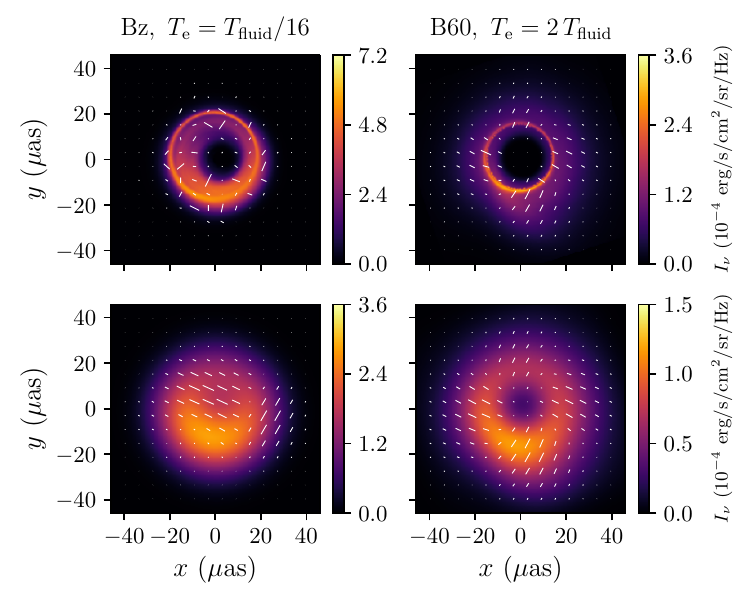}
    \caption{Same as Fig.~\ref{fig:polar-img-0}, but for two different cases with higher and lower electron temperature models that give an image orientation $\eta<180^{\circ}$, more consistent with the EHT M87 data. In the left column, $T_e=T_{\mathrm{fluid}}/16$ and there is significant emission in the foreground that changes the image orientation. In the right column, $T_e=2\,T_{\mathrm{fluid}}$ and the change in image orientation is due to the emission at larger radii.}
    \label{fig:polar-img-1}
\end{figure}

As expected, the images of moderate electron temperature models (e.g. Fig.~\ref{fig:polar-img-0} with $T_{\mathrm{e}}=T_{\mathrm{fluid}}/2$) are dominated by a ring with a diameter of $\simeq 30-40\,\mu$as, whose asymmetry is mostly caused by Doppler beaming.  The latter implies a relatively robust orientation of the asymmetric image (quantified by $\eta$ in Fig.~\ref{fig:ring-reports-0}). This is less true for higher or lower electron temperature models. In particular, for higher (lower) temperature models, the disk (jet) can become somewhat more prominent outside (inside) the ring (as shown with the two examples in Fig.~\ref{fig:polar-img-1}). This is responsible for the increase (decrease) of the ring diameter, and the decrease (increase) of the fractional central brightness seen in the ring statistics in Fig.~\ref{fig:ring-reports-0} with increasing (decreasing) $T_e$. The jet and outer disk can also have additional asymmetry due to the non-uniform distribution of the luminous plasma, where the intensity can peak at different locations in the image, depending on the detailed fluid properties of the simulation. This explains why in Fig.~\ref{fig:ring-reports-0}, for higher and lower electron temperature models, the dispersion of the orientation angle $\eta$ becomes larger, while the overall asymmetry $A$ tends to decrease, as the contributions from different radii may cancel out.

Assuming a $288^{\circ}$ jet position angle \citep{2018ApJ...855..128W}, the asymmetry from the main image ring alone would lead to an image orientation angle of $\eta>180^{\circ}$, while this quantity is reported to be smaller than $180^{\circ}$ in EHT measurements \citep{2019ApJ...875L...3E}. This tension is interpreted as a hint of tilted gas angular momentum in \citet{2020MNRAS.499..362C}.  However, as shown in Fig.~\ref{fig:polar-img-1}, the additional asymmetry introduced by the jet and/or the outer disk may also lead to a smaller $\eta$, providing an alternative remedy for the observed tension. We find that this effect is more prominent in our spherical MAD simulations relative to the rotating initial conditions considered in previous EHT modeling. Therefore, more follow-up data and theoretical modeling are required to determine the implication of the small $\eta$ value measured by EHT.

To summarize, Fig.~\ref{fig:ring-reports-0} shows that the intensity images are overall very similar among the different GRMHD simulations considered in this paper. The viability of using intensity images to distinguish between different magnetic tilts is largely limited by the uncertainties in e.g. the electron temperature models.

\section{Spectra of Synchrotron Emission}

\label{sec:spectra}

\begin{figure}
	\centering
    \includegraphics[width=\columnwidth]{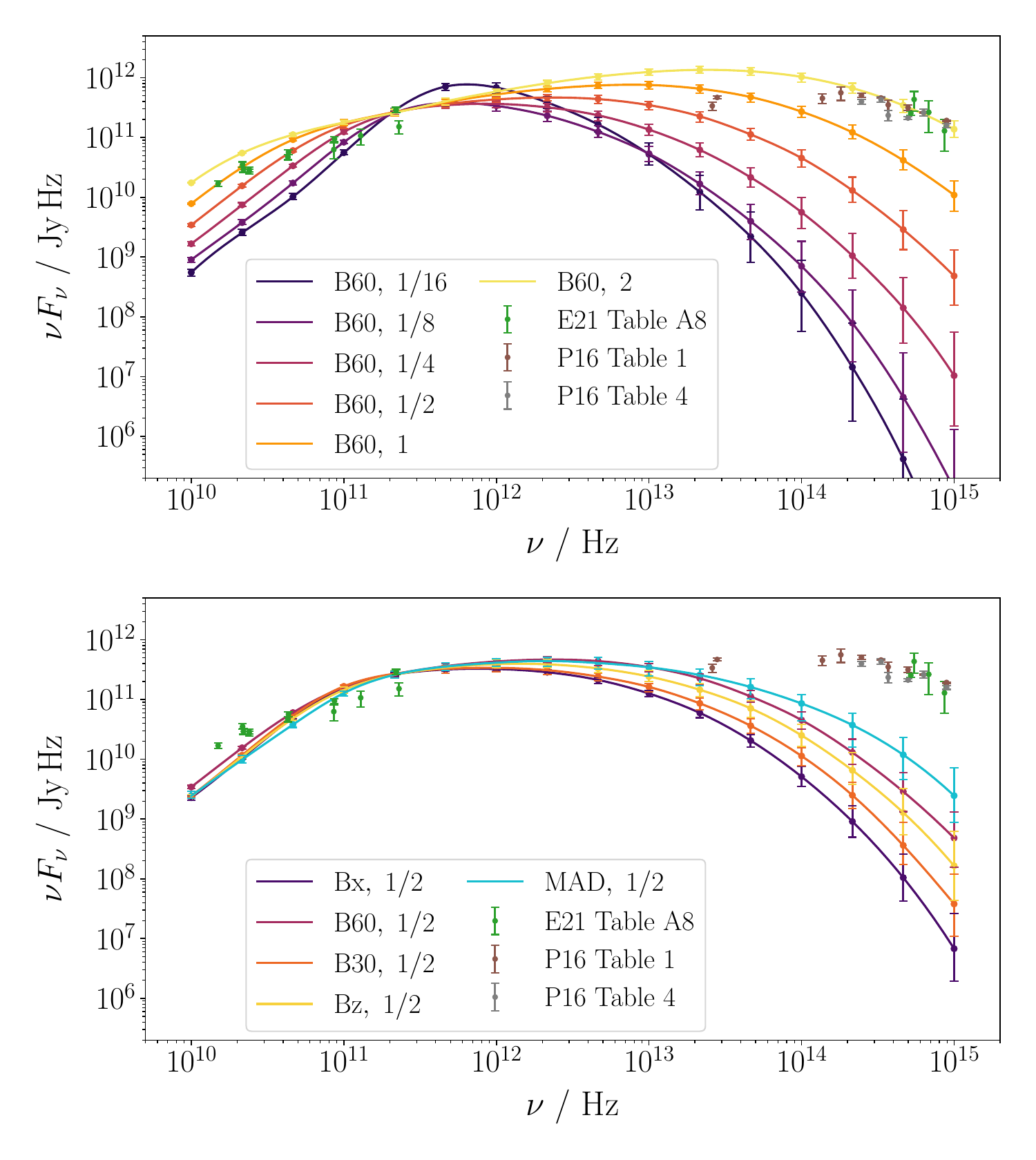}
    \caption{Synchrotron emission spectra evaluated by ray tracing at 16 frequencies from radio to UV. The labels stand for the corresponding GRMHD simulations and electron temperature models, so for example, "B60, 1/16" means the B60 simulation with $T_e=T_{\mathrm{fluid}}/16$. For each model, we show the median of the intensity flux density over the 144 (100 for MAD) snapshots, with error bars given by the 25\% and 75\% percentiles. In the top panel, we focus on the B60 simulation and compare different electron temperature models, while in the bottom panel, we fix $T_e=T_{\mathrm{fluid}}/2$ and compare different GRMHD simulations. Observed data for M87 are taken from \citet{2021ApJ...911L..11E} and \citet{2016MNRAS.457.3801P}. The synchrotron emission spectra of different GRMHD simulations are very similar except at higher frequencies, where a stronger large scale jet leads to a larger synchrotron emission flux.}
    \label{fig:spectra}
\end{figure}

To constrain the emission as a function of frequency in our models, we use the EHT Multi-Wavelength Science Working Group's series of broad-band spectrum measurements for M87, which were quasi-simultaneous to the EHT M87 imaging campaign \citep{2021ApJ...911L..11E}. The radio angular resolution is approximately $10^{-3}$ arcsec in that work. We also include the infrared and visible band data from \citet{2016MNRAS.457.3801P}, for which the aperture radius is about 0.1-0.4 arcsec.

For each model, we run \texttt{grtrans} ray tracing at 16 different frequencies, evenly spaced in log space between $10^{10}$ and $10^{15}$ Hz, with a resolution of $128^2$ pixels. We use 144 snapshots spaced evenly over 14.6 years ($14{,}400\,GM/c^3$) for the spherical simulations, and 100 snapshots spaced evenly over 4.1 years ($4000\,GM/c^3$) for MAD simulations. The overall normalization of density $[\rho]$ is still calibrated such that the flux density at 230 GHz is 1.2 Jy. The resulting spectra are presented in Fig.~\ref{fig:spectra}. Note that we do not include inverse Compton emission which may become important at higher frequencies (e.g., \citealt{Ryan2018}).

We compare the different simulations while fixing $T_e=T_{\mathrm{fluid}}/2$ in the bottom panel of Fig.~\ref{fig:spectra}. Although all of the models are adjusted to have the same flux at 230$\,$GHz, it is nonetheless remarkable how similar the spectra are from $10^{10}$ to $10^{13}\,$Hz; one needs to go to even higher frequencies to see a difference. Interestingly, the ranking of the high frequency flux in the bottom panel of Fig.~\ref{fig:spectra} coincides with the ranking of jet strength illustrated in Fig.~\ref{fig:volume-rendering}. Since the characteristic synchrotron frequency $\nu_s$ is proportional to $T_e^2$, at higher frequencies the synchrotron emission is dominated by the large $T_e$ regions, which corresponds to the jet (see the $T_{\mathrm{fluid}}$ plots in Fig.~\ref{fig:nTbeta-0}). This explains why a stronger jet leads to a larger high frequency flux, given the thermodynamics in our GRMHD simulations.

The top panel of Fig.~\ref{fig:spectra} compares the spectra of the B60 simulation with different $T_e$ models. As expected, the frequency of peak intensity increases with increasing $T_e/T_{\mathrm{fluid}}$ since $\nu_s\propto T_e^2$. The spectrum is significantly flatter for larger $T_e/T_{\mathrm{fluid}}$, at both high and low frequency side. This trend is similar for other GRMHD simulations. We note that since we only include synchrotron emission in ray tracing, here the  measurements should be interpreted as upper bounds on the model predictions, particularly at higher frequencies. In this sense, the low electron temperature models are mostly consistent with the observations, while there is some tension for the highest electron temperature models such as $T_e=2\,T_{\mathrm{fluid}}$. Such tension may however be affected by the uncertainty in the overall normalization of the spectra: around 230 GHz, the data points in \citet{2021ApJ...911L..11E} show large dispersion. While the higher value reflects the total flux density from the center of M87, the lower value employs a core-jet model which removes the contribution from the extended \textit{jet} and only contains the portion from the M87 \textit{core}. In this paper, we adopt the value of 1.2 Jy based on the total flux density measured on April 11, 2017 \citep[][Table 4]{2019ApJ...875L...3E}, and do not include any core-jet model to avoid the additional uncertainties it may introduce. Had we calibrated the flux density at 230 GHz to a smaller value, the tension of the radio and infrared band spectra for higher electron temperature models in Fig.~\ref{fig:spectra} would be relieved.

\begin{figure}
	\includegraphics[height=20.3cm]{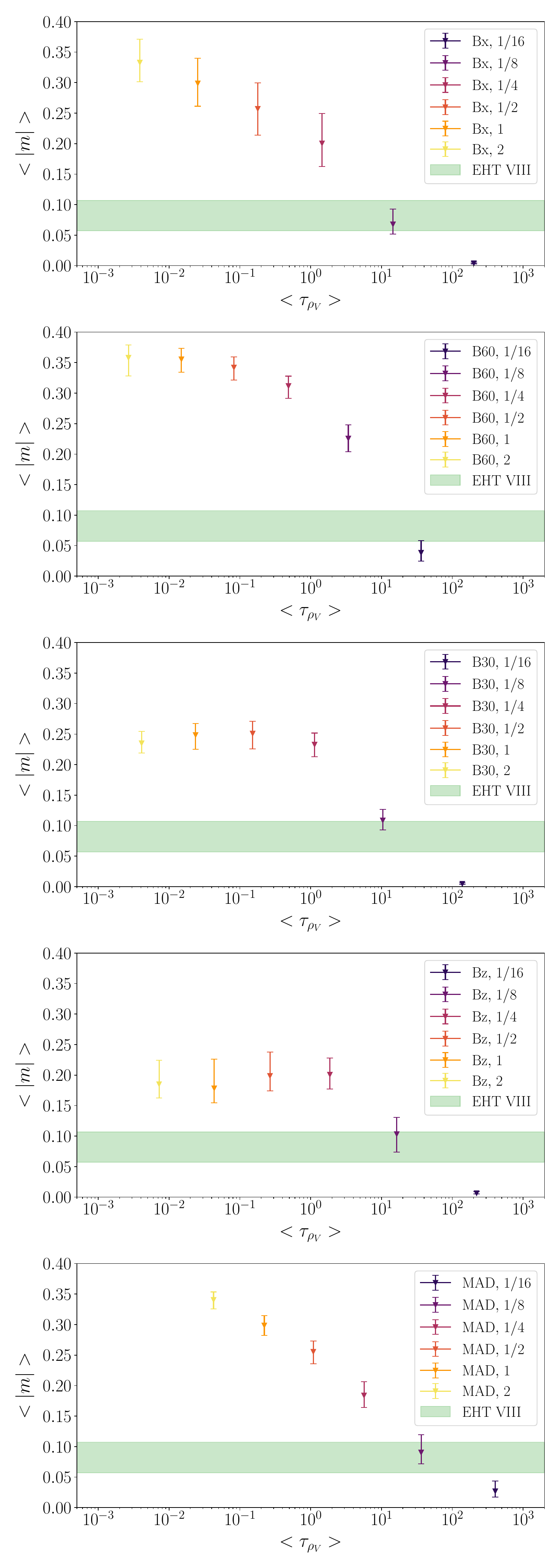}
    \caption{The pixel-level fractional polarization $\left<|m|\right>$, as a function of the intensity-weighted mean Faraday rotation depth, $\left<\tau_{\rho_V}\right>$. The values are reported in terms of the 25\%, 50\% and 75\% percentiles. For lower electron temperature models, a larger accretion rate is required to produce the same flux at 230$\,$GHz, leading to a larger Faraday rotation depth. The green band indicates the EHT measurement $5.7\% \leq \left<|m|\right> \leq 10.7\%$ \citep{2021ApJ...910L..13E}. Our Bz model shows significantly lower factional polarization at low Faraday depth than the other models, but still not quite low enough for consistency with the EHT data.}
    \label{fig:m-0}
\end{figure}

\section{Polarimetric Signatures at 230 GHz}

\label{sec:polar}

\begin{figure*}
	\includegraphics[width=\textwidth]{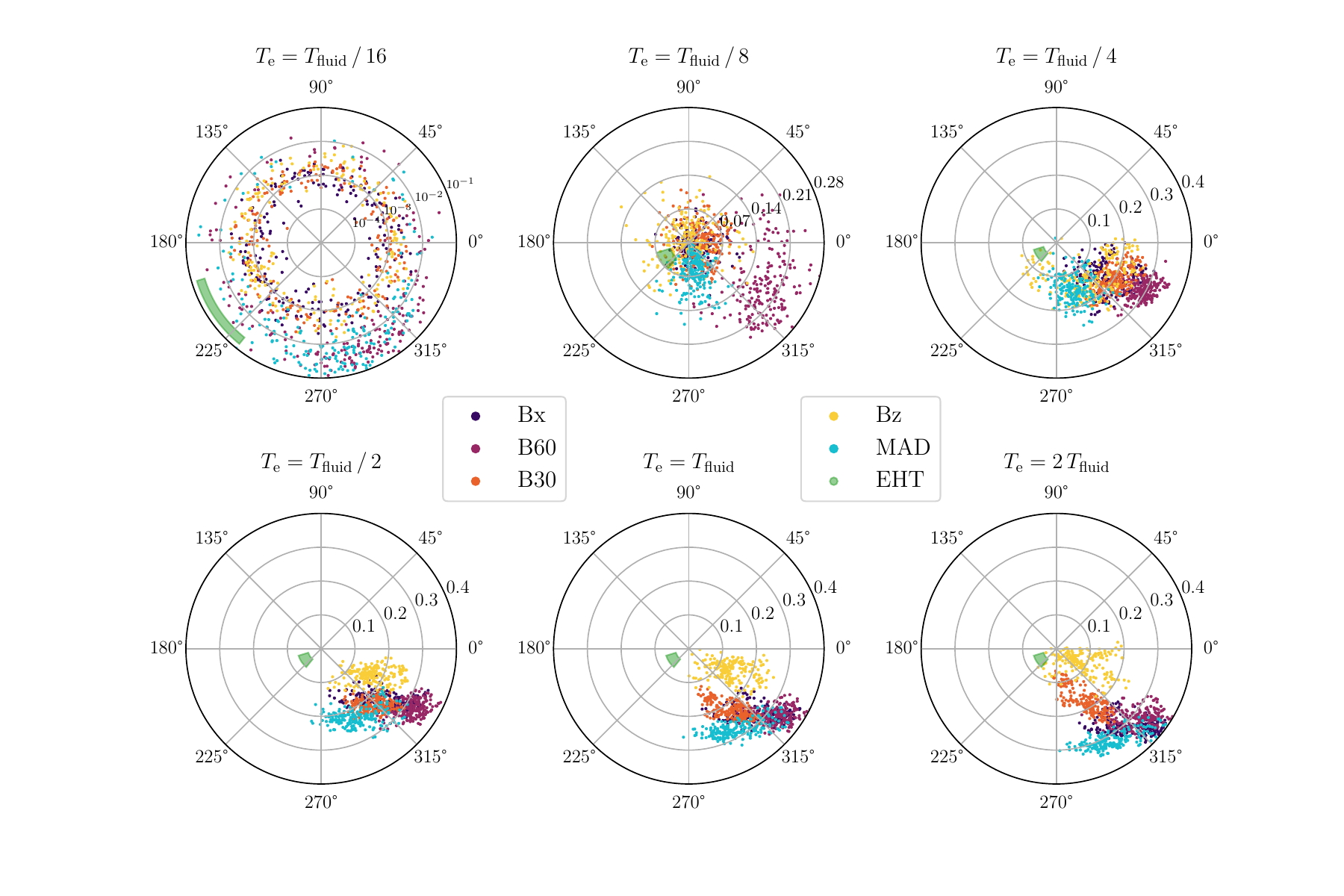}
    \caption{Polarization of our horizon-scale images analyzed using the $\beta_2$ statistics introduced by \citet{2021ApJ...911L..11E}. The modulus of $\beta_2$ characterizes the fraction of the azimuthally symmetric polarized component, while the phase indicates the specific polarization pattern. Each dot represents a snapshot. For low electron temperature models, due to Faraday depolarization, $|\beta_2|$ is small and $\arg [\beta_2]$ is random. For high electron temperature models, $\arg [\beta_2]$ fluctuates around $315^{\circ}$ because in these GRMHD simulations, the magnetic field is more azimuthal than radial in the region where most emission comes from. The green partial wedge indicates the EHT measurements $0.04 \leq |\beta_2| \leq 0.07$ and $197^{\circ} \leq \arg [\beta_2] \leq 231^{\circ}$ \citep{2021ApJ...910L..13E}. The horizon-scale polarization properties are similar across all the simulations, and insensitive to the initial magnetic field orientation and the presence/absence of initial rotation of the gas.}
    \label{fig:beta2-0}
\end{figure*}

\begin{figure*}
	\includegraphics[width=\textwidth]{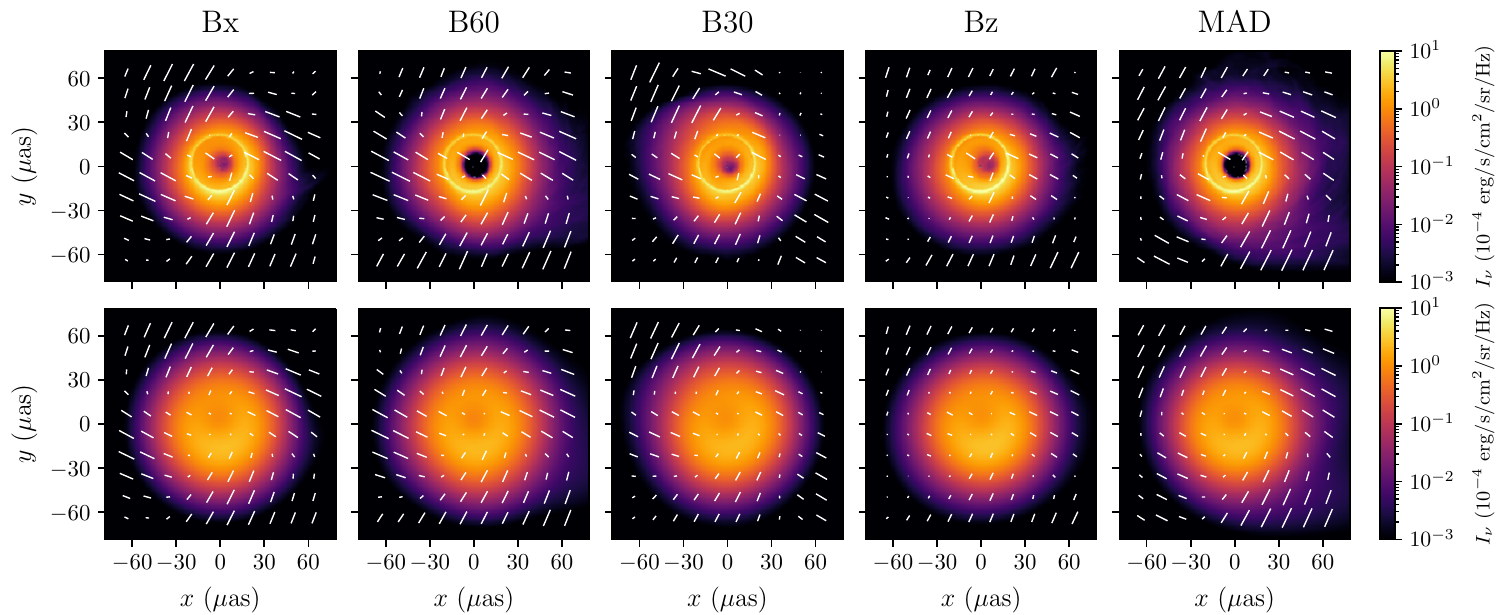}
    \caption{Same as Fig.~\ref{fig:polar-img-0}, but with a logarithmic color scale to display a larger dynamic range for the image. Note that here the tick length is proportional to the pixel-level fractional linear polarization $\mathcal{P}/\mathcal{I}$, as opposed to Figs.~\ref{fig:polar-img-0} and~\ref{fig:polar-img-1} where it represents the linearly polarized flux $\mathcal{P}$. With the same electron temperature model $T_e=T_{\mathrm{fluid}}/2$, Bz has the overall smallest fractional polarization, similar to Fig.~\ref{fig:m-0}. However, the polarization direction is similar among the different simulations out to $60\,\mu$as ($15.7\,r_g$ for M87), which corresponds to an image dynamic range of $\sim\!10^4$, far beyond the planned $\sim\!10^2$ dynamic range of ngEHT \citep{2019BAAS...51g.256D}.}
    \label{fig:polar-img-2}
\end{figure*}

To compare the simulated polarimetric images with EHT observations, we compute the $\left<|m|\right>$ and $\beta_2$ statistics, as defined in Equations \ref{eq:m} and \ref{eq:beta_m}. In Fig.~\ref{fig:m-0}, we plot $\left<|m|\right>$ as a function of the intensity-weighted mean Faraday rotation depth $\left<\tau_{\rho_V}\right>$. Generally, lower electron temperature models need a larger accretion rate to match the observed flux at 230$\,$GHz, which will then lead to a larger $\left<\tau_{\rho_V}\right>$. When $\left<\tau_{\rho_V}\right> \gtrsim 1$, the effects of Faraday depolarization will be significant, so that $\left<|m|\right>$ decreases with increasing $\left<\tau_{\rho_V}\right>$ for all the simulations. When $\left<\tau_{\rho_V}\right> \ll 1$, Faraday depolarization can be neglected, and the different electron temperature models mainly affect $\left<|m|\right>$ by changing the region where most emission comes from. For the B30, B60 and Bz simulations, we see a point beyond which $\left<|m|\right>$ no longer increases with increasing $T_e/T_{\mathrm{fluid}}$, while for Bx and MAD, such a turning point does not appear. The EHT measurement suggests that $5.7\% \leq \left<|m|\right> \leq 10.7\%$ for M87, which would favor $T_e \sim T_{\mathrm{fluid}}/8$. However, it is important to note that there is a factor of 2 difference in $\left<|m|\right>$ at high $T_e$ (low $\left<\tau_{\rho_V}\right>$) among the different GRMHD simulations. It is plausible that yet different initial conditions (e.g. a different Bondi radius or initial magnetic field structure) could further decrease the value of $\left<|m|\right>$. Therefore, we believe that it is premature to exclude the higher $T_e$ models based on the EHT $\left<|m|\right>$ measurement.

While $\left<|m|\right>$ indicates the pixel-level degree of polarization, the $\beta_2$ statistics also contain information about the pattern of the azimuthally symmetric polarised component. For example, $\beta_2=1\,(-1)$ stands for purely radial (azimuthal) polarization, and $\beta_2=\pm\,i$ represents the intermediate states where the radial and azimuthal components are equal \citep[see e.g. Fig.~20 of][]{2021ApJ...910L..13E}. In Fig.~\ref{fig:beta2-0}, we show the scatter plot of $\beta_2$ for different GRMHD simulations and electron temperature models, using 200 snapshots for each combination. Similar to Fig.~\ref{fig:m-0}, here for the lowest electron temperature model $T_e=T_{\mathrm{fluid}}/16$, the overwhelming Faraday depolarization leads to a tiny $|\beta_2|$ and an almost random $\arg[\beta_2]$. As we increase $T_e/T_{\mathrm{fluid}}$, all the GRMHD simulations begin to favor $\arg[\beta_2] \sim 315^{\circ}$, which is in tension with the EHT measurement $197^{\circ} \leq \arg [\beta_2] \leq 231^{\circ}$. As the polarization direction is by definition perpendicular to the magnetic field direction, here the fact that the simulated polarization pattern is too radial indicates that the magnetic field is probably too azimuthal \citep{2021ApJ...912...35N}.

One striking result in Fig.~\ref{fig:beta2-0} is that the polarization orientation $\arg[\beta_2]$ is not sensitive to the initial magnetic field tilt. Roughly speaking, the magnetic field can be considered as a superposition of the initial field (which is asymptotically constant at larger radii) plus the field induced by the inflow in the accretion flow (which should vanish at very large radii but dominate at smaller radii). Therefore, one might think that if we could measure the polarization pattern at larger radii, we would be able to recover the magnetic tilt information. However, in Fig.~\ref{fig:polar-img-2}, which show the same images as Fig.~\ref{fig:polar-img-0} but on a logarithmic scale; we find that the polarization direction is similar among our different GRMHD simulations out to $60\,\mu$as ($15.7\,r_g$ for M87). Since the 230$\,$GHz intensity at $60\,\mu$as is about $10^4$ times smaller than the brightest spot in the inner ring, this is already beyond the reach of the planned next-generation EHT \citep[ngEHT,][]{2019BAAS...51g.256D} with a targeted image dynamic range of $\sim\!10^2$.

Even starting with very different initial magnetic field orientations at $r\sim 10^2\,r_g$, we find that on the spatial scales accessible with future EHT observations, the polarization of the image does not depend that strongly on the initial magnetic field orientation. Perhaps most interestingly, even the case with no net magnetic flux along the spin axis (our Bx simulation) produces a similar polarization pattern both on horizon scales and out to $\sim 10 - 20 \,r_g$ (see Figs.~\ref{fig:polar-img-0} and~\ref{fig:polar-img-2}). This demonstrates that the local geometry of the magnetic field is shaped in our spherical simulations by a combination of inflow, reconnection and frame dragging; these are more important for the polarization of the synchrotron emission than the initial net flux. The emission at larger distances might be expected to contain more information about the magnetic field supplied to the flow from large radii. This larger-scale emission may, however, be dominated by the jet sheath, which is more difficult to predict \citep[e.g.,][]{2019MNRAS.486.2873C} and is strongly influenced, of course, by the near horizon magnetic field structure. We find that for resolutions and dynamic range achievable in the near-future, even the emission at larger radii far from the jet does not contain much information about the  unperturbed field being advected in from large radii (Fig.~\ref{fig:polar-img-2}).

\section{Discussion}

\label{sec:discussion}

In this paper, we have explored the observational signatures of GRMHD magnetically arrested (MAD) simulations of initially non-rotating matter accreting onto a rotating black hole (see \citealt{2021MNRAS.504.6076R}). The key parameter that varies across the different simulations is the initial tilt of the magnetic field relative to the black hole spin axis (which varies  from a magnetic field that is fully aligned with the spin to one that is orthogonal to it).  We have also compared the results of these simulations to more standard initial conditions with a rotationally supported torus (e.g., \citealt{2011MNRAS.418L..79T}).  Part of our motivation is to assess the extent to which the horizon-scale dynamics and observational signatures are sensitive to the uncertain nature of the inflowing plasma at larger distances from the black hole.   We have focused on quantifying the mm images and polarization patterns, as would be observed by the EHT, as well as the synchrotron spectra from the radio to optical-IR.

We find distinct similarity among the different GRMHD simulations in the intensity images (Section \ref{sec:intensity}), spectral energy distributions (Section \ref{sec:spectra}) and polarimetric signatures (Section \ref{sec:polar}).   \citet{2021MNRAS.504.6076R} quantified the dynamical similarity of these different simulations, and their similarity to rotationally supported MADs (see Figs.~\ref{fig:nTbeta-0} and~\ref{fig:nTbeta-1}).  Here we have shown that this similarity extends to EHT observables. Indeed, the quantitative similarity in the images (Figs.~\ref{fig:polar-img-0}, \ref{fig:ring-reports-0}, and~\ref{fig:polar-img-2}), spectra (Fig.~\ref{fig:spectra}), and polarization (Figs.~\ref{fig:polar-img-0}, \ref{fig:beta2-0}, and~\ref{fig:polar-img-2}) among the different simulations is remarkable, even comparing the simulation with no net magnetic flux ("Bx") in the direction of the black hole spin to the simulation in which the initial magnetic field is fully aligned with the BH spin ("Bz").  We have also considered future more sensitive EHT-like experiments (e.g., \citealt{2019BAAS...51g.256D}) which will plausibly have a factor of $\sim 100$ in dynamic range in intensity and polarization.   Even with such increased dynamic range, we find that the images and polarization patterns are very similar for the full set of simulations we consider (Fig.~\ref{fig:polar-img-2}). Although this paper mainly focuses on the observational signatures of M87 (with viewing angle $\theta_0=163^{\circ}$), we also checked the images for $\theta_0=135^{\circ}$. In that case, the black hole images are more asymmetric due to stronger Doppler beaming, but the comparison of the different simulations remains similar; this suggests that the conclusions of this paper may also be generalized to other black hole accretion systems.

Heuristically, the natural interpretation of these results is that close to the black hole the initial, uniform magnetic field becomes comparatively unimportant compared with the magnetic field induced by the accretion flow and the spin of the central black hole.  For many of the dynamical and observational properties of our simulations, it is not even critical how much magnetic flux is supplied along the spin axis of the black hole (see below for the key exception related to the strength of the jet).  It is, however, important that the initial magnetic field is relatively strong and ordered so that the different simulations are all roughly magnetically arrested.   

In some of the electron temperature models considered in this work, the jet in the foreground and/or the plasma at large radii contribute significantly to the observed emission, changing the image morphology; examples are shown in Fig.~\ref{fig:polar-img-1}. This effect is particularly prominent in low and high electron temperature models and is more pronounced in the simulations with non-rotating initial conditions than in the simulations with rotating initial conditions.   One consequence of this is that the orientation of the bright region in the images of our mock EHT M87 observations can be $\eta < 180^{\circ}$, consistent with EHT measurements \citep{2019ApJ...875L...3E}.  Typical simulations with initially rotating MADs do not reproduce this image orientation, though \citet{2020MNRAS.499..362C} showed that the image orientation could be explained if the gas angular momentum is tilted relative to the black hole angular momentum.   More work is clearly required to determine the physical origin of the image orientation in EHT observations of M87.   

Large spin MAD models for M87 are favored by jet power measurements (e.g., \citealt{2006MNRAS.370..981S}).   However, they do not produce the correct value of the polarization parameter  $\beta_2$ introduced by the EHT (Fig.~\ref{fig:beta2-0}), which measures the orientation of the polarization pattern and is in turn determined by the structure of the magnetic field in the emitting region (e.g., how radial or azimuthal the field is).   We find that this discrepancy persists for all of the  electron temperature and magnetic tilt models studied in this work.   This tension is remarkable, as it is more robust against the large uncertainties in the electron temperature models, compared with e.g. the properties of the ring in the intensity image (Fig.~\ref{fig:ring-reports-0}) or the pixel-level fractional polarization $\left<|m|\right>$ (Fig.~\ref{fig:m-0}). It would be interesting to explore what is required to make the horizon scale magnetic field less azimuthal in such MAD models so as to match the EHT polarization measurements.

We have explored a variety of electron temperature models in this work, and find that the similarity among the different MAD models is relatively robust to the precise electron temperature prescription we consider (e.g., Figs.~\ref{fig:ring-reports-0} and~\ref{fig:beta2-0}). This does not guarantee, of course, that other electron temperature models (e.g., ones with emission at only high $\sigma$) or other observational diagnostics (e.g., application to higher accretion rate systems with prominent inverse Compton emission) will not reveal more significant differences.   

There is one key property of the simulations we have considered that {\em does} depend significantly on the rotation and magnetic field structure supplied at large radii: the strength and stability of the resulting jet.    \citet{2021MNRAS.504.6076R} (see their Fig.~8) showed that the dimensionless magnetic flux threading the horizon varied from $\phi \simeq 20-80$ for the different magnetic field orientations considered here (with "B60" having the largest flux on the horizon and "Bx" the smallest) and that there are correspondingly a factor of $\sim 30$ differences in the jet power among these different simulations.  Even the simulation with no initial magnetic flux along the spin axis of the BH ("Bx") does transiently produce a significant jet (albeit the weakest among the simulations we consider), due to vertical magnetic field created by reconnection-driven turbulence in the inflow near the black hole.  There are constraints on the power of the jet in M87, but the uncertainties are large: current results range from $10^{42}\,\mathrm{erg}\,\mathrm{s}^{-1}$ to $10^{45}\,\mathrm{erg}\,\mathrm{s}^{-1}$ \citep[e.g.][]{2006MNRAS.370..981S,2015ApJ...805..179B,2016MNRAS.457.3801P}.   The observational uncertainty in the jet power is thus larger than the theoretical uncertainty associated with the different versions of MADs considered in this paper.   

Correspondingly, Fig.~\ref{fig:volume-rendering} shows volume renderings of the jets in these different simulations.   The initially rotating MAD produces a strong, stable jet, while all of the initially non-rotating MADs produce jets that undergo the kink instability at radii $\gtrsim 100\,r_g$, triggered by the confining pressure of the ambient spherical inflow.  It is very likely that observational signatures of the jet at low frequencies in the radio would show significant differences due to these different stability properties at large radii.   Modeling such emission requires, however, a treatment of non-thermal electron acceleration in the jet at large distances from the black hole $\sim 10^2-10^3 \,r_g$.   We leave this to future work. 

In contrast to the relatively weak dependence of our results on the tilt of the magnetic field relative to the black hole spin axis, analogous calculations of rotating flows with the angular momentum tilted relative to the black hole spin axis find that the fluid properties are quite different at radii $\lesssim 10\,r_g$ (e.g., \citealt{2007ApJ...668..417F,2013Sci...339...49M}). This leads to significant differences in EHT synchrotron emission observables as a function of angular momentum tilt \citep{2020ApJ...894...14W,2020MNRAS.499..362C}. Our combined results thus suggest that horizon-scale observations are more sensitive to the tilt of the angular momentum of the inflowing gas than the tilt of the inflowing magnetic field.

\section*{Acknowledgements}

We thank Andrew Chael, Alex Lupsasca, and George Wong for helpful discussions.  This work was supported in part by a Simons Investigator award from the Simons Foundation (EQ) and by NSF AST 1715054.  SMR was supported by the Gordon and Betty Moore Foundation through Grant GBMF7392.  This research was supported in part by the National Science Foundation (NSF) under Grant No. NSF PHY-1748958, and by the NSF through XSEDE computational time allocation TG-AST200005 on Stampede2. The initial simulations for this work were made possible by computing time granted by UCB on the Savio cluster.   The analysis presented in this article was performed on computational resources managed and supported by Princeton Research Computing, a consortium of groups including the Princeton Institute for Computational Science and Engineering (PICSciE) and the Office of Information Technology's High Performance Computing Center and Visualization Laboratory at Princeton University.

\section*{Data Availability}

The data underlying this paper will be shared on reasonable request
to the corresponding author.



\bibliographystyle{mnras}
\bibliography{example} 


\bsp	
\label{lastpage}
\end{document}